**An organic-nanoparticle transistor behaving as a biological spiking synapse ****


By *Fabien Alibart*, *Stéphane Pleutin*, *David Guérin*, *Christophe Novembre*, *Stéphane Lenfant*, *Kamal Lmimouni*, *Christian Gamrat* and *Dominique Vuillaume**

[*]     Dr. F. Alibart, Dr. S. Pleutin, Dr. D. Guérin, Dr. S. Lenfant, Dr. K. Lmimouni, Dr. D. Vuillaume
Molecular Nanostructures and Devices group, Institute for Electronics Microelectronics and Nanotechnology (IEMN), CNRS, University of Lille, BP60069, avenue Poincaré, F-59652cedex, Villeneuve d'Ascq (France).
E-mail: dominique.vuillaume@iemn.univ-lille1.fr
        Dr. C. Novembre, Dr. C. Gamrat
CEA, LIST/LCE (Advanced Computer technologies and Architectures), Bat. 528, F-91191, Gif-sur-Yvette (France).



[**]    This work was funded by the European Union through the FP7 Project NABAB (Contract FP7-216777) and partly by the Micro and Nanotechnology Program from the French Ministry of Research under the grant RTB: "Post CMOS moléculaire". We thank R. Baptist (CEA-LETI) for his support. Supporting Information is available online from Wiley InterScience or from the corresponding author.

Molecule-based devices are envisioned to complement silicon devices by providing new

functions or already existing functions at a simpler process level and at a lower cost by virtue

of their self-organization capabilities. Moreover, they are not bound to von Neuman

architecture and this feature may open the way to other architectural paradigms.

Neuromorphic electronics is one of them. Here we demonstrate a device made of molecules

and nanoparticles, a nanoparticle organic memory filed-effect transistor (NOMFET), which

exhibits the main behavior of a biological spiking synapse. Facilitating and depressing

synaptic behaviors can be reproduced by the NOMFET and can be programmed. The synaptic

plasticity for real time computing is evidenced and described by a simple model. These results





open the way to rate coding utilization of the NOMFET in dynamical neuromorphic computing circuits.

## 1. Introduction

It is now well recognized that electronic circuits based on von Neuman paradigm are unable to catch the complex, real-world environment, behaviors as do a biological neural systems (e.g. human brain). One of the reason is the so-called von Neuman bottleneck[1] due to the physical separation of computing units and memories. In the brain, memory and computation are mixed together and allow the processing of information both in time and in space via the time dependent properties of interconnected neurons. [1,2] This challenge for the development of a new generation of computers has induced a lot of efforts in neuroscience computation activities and the framework for the spatiotemporal processing of information seems to be theoretically achievable.[2] One key element that is still a limitation concerns the integration of neurons and synaptic connections in order to realize a brain-like computer. Even if silicon CMOS chips have been designed and fabricated to emulate the brain behaviors,[3, 4] this approach is limited to small system because it takes several (at least 7) silicon transistors to build an electronic synapse.[5] As the human brain contain more synapses than neurons (by a factor of ~ $10^7$), it is mandatory to develop a nano-scale, low power, synapse-like device if we want to scale neuromorphic circuits towards the human brain level. This feature has recently prompted the research for nano-scale synaptic devices. Proposals for such programmable memory devices include optically-gated CNTFET,[6,7] organic/hybrid Si nanowire transistor,[8] memristor.[9-12]

Here, we show that mixing nanoparticles (NPs) and molecules to implement computation and memory in a single synapse-like device is a powerful approach towards such objectives. NPs and molecules are nano-size objects suitable for nano-device fabrication, they can be manipulated and assembled by low-cost, bottom-up, techniques (e.g. self-assembly),[13,





[14] they are prone to work on flexible, plastic, substrates (see the tremendous efforts on plastic, printable, organic electronics[15-17]).

The nanoparticle organic memory field-effect transistor (NOMFET) demonstrated here can be programmed to work as a facilitating or depressing synapse; it exhibits short-term plasticity (STP) for dynamical processing of spikes. This behavior is obtained by virtue of the combination of two properties of the NOMFET: the transconductance gain of the transistor and the memory effect due to charges stored in the NPs. The NPs are used as nanoscale capacitance to store the electrical charges and they are embedded into an organic semiconductor (pentacene). Thus, as detailed in section 4, the transconductance of the transistor can be dynamically tuned by the amount of charges in the NPs. We previously demonstrated that this type of device works as a memory[18] but with a "leaky" behavior. The retention times are in the range of few seconds to few $10^3$ s. This behavior is used here to implement the synaptic weight $w_{ij}$ with a possible dynamic working in this time range, a mandatory condition to obtain the training/learning capabilities of a spiking neural network.[19] A transistor is basically a multiplier, it is used to realize the basic function of the synapse described as $S_j=w_{ij}S_i$, where $S_i$ and $S_j$ are the pre- and post-synaptic signals (here the source/drain current and voltage of the NOMFET). The synaptic weight $w_{ij}$ is a time-dependent parameter whose value depends on the activity of the pre- and post-synapse neurons. We demonstrate that we can tailor the dynamic behavior of the NOMFET in the frequency/time domain (0.01-10 Hz) by adjusting the size of both the NPs (5-20 nm in diameter) and the size of the NOMFET (channel length L from 200 nm to 20 μm). We also demonstrate that models developed to explain and simulate the plasticity of biological synapses can be successfully adapted to the NOMFET behavior. These results open the way to the rate coding utilization[20] of the NOMFET in neuromorphic computing circuits.[21]





## 2. NOMFET – materials and fabrication.

The NOMFET (Fig. 1) consists of a bottom-gate, bottom source-drain contacts organic transistor configuration. We immobilized the gold NPs (20, 10 and 5 nm in diameter) into the source-drain channel using surface chemistry (self-assembled monolayers or SAM). They were subsequently covered by a thin film (35 nm thick) of pentacene (see experimental section). This device gathers the behavior of a transistor and a memory.[18] We selected several sizes for the NPs and channel lengths of the NOMFET (200 nm to 20 $\mu$m) to study their impact on the synaptic behavior of the NOMFET. We also studied the dependence of this behavior as a function of the NP density. Figure 1 shows scanning electron microscope (SEM) images of NP arrays with NP densities ranging from ~$10^{10}$ cm$^{-2}$ to a quasi-continuous 2D film. The density of the NP network is controlled by the density of NPs in solution and the duration of the reaction (see experimental section). The optimum NP density to observe the synaptic behavior reported here is between $10^{11}$ and $10^{12}$ cm$^{-2}$ as shown in Figs. 1-b to 1-d for the 5 nm, 10 nm and 20 nm NPs, respectively. A too low density, <$10^{11}$ NP/cm$^2$, (fig. 1-a) leads to a too weak memory and synaptic behaviors of the NOMFET. At high density, >$10^{12}$ NP/cm$^2$, (above the percolation threshold, such as in fig. 1-e), the device does not exhibit a transistor behavior due to metallic shorts between source and drain, and screening of the gate voltage by the metallic film of NPs. In the following, we focus on devices with a NP density of around $10^{11}$ cm$^{-2}$. In all cases (figs 1-b to 1-d), we obtained a rather uniform distribution of NPs (no NP aggregation) with a density of ~ $10^{11}$ NP/cm$^2$. Even if we have optimized the process (see experimental section) to got a reproducibility (i.e. a density of about $10^{11}$ NP/cm$^2$ without aggregation) of nearly 100% in the case of the 10 and 20 nm NPs (tested on three different fabrication runs), we have observed in few cases the formation of some aggregates (see supporting information, fig. S3). A reproducibility of only 50% is obtained for the 5 nm NPs, mainly due the lack of reproducibility to form the thiol-terminated SAM on the gate





oxide (see experimetal section). Noteworthy, we always observed the synaptic behavior of the NOMFET, even when the arrays of NPs seem less under controlled (see section 4). This feature means that the synaptic behavior of the NOMFET is very robust and defect-tolerant, and this feature can be viewed as an important advantage for the envisioned applications in neuronal computing at the nanoscale.

We checked by TM-AFM the morphology of the pentacene layer deposited over the NP networks. Figure 2 shows the pentacene morphology for the reference sample (no NPs) and for the 10 nm NP NOMFET. The film without NPs shows the usual polycrystalline structure of pentacene film with large grain size and terraces, each terrace corresponding to a monolayer of pentacene molecules. [22] The pentacene film is more disordered in the presence of the NPs with a smaller grain size, and the terraces are more difficult to see. This result may be due to both the presence of NPs that hinders the surface diffusion of pentacene molecules during the deposition, and the presence of the organic SAM used to anchor the NPs on the surface. Smaller grain size has been often observed for OFET with a gate dielectric functionalized by a SAM. [23] This more disordered pentacene film explains the decrease in the output drain current $I_D$ and hole mobility $\mu_h$ measured for the NOMFET ($\sim 10^{-3}$ cm$^2$V$^{-1}$s$^{-1}$) compared to the reference device ($\sim 0.1$ cm$^2$V$^{-1}$s$^{-1}$) – see table 1. These values are averaged over 6-8 devices, and we notice that the presence of NPs induces a larger dispersion of the FET parameters. The error bars given in table 1 are the standard deviation of a normal distribution fitted on the data. However, we do not observe a specific trend as function of the NP size (table 1).

## 3. Basic behavior of a biological synapse

For the sake of clarity and comparison with the NOMFET, let us start with a brief description of how a biological synapse works. The most important feature of a synapse is its ability to transmit in a given way an action potential (AP) from one pre-synapse neuron N1, to a post-





synapse neuron N2. When a sequence of APs is sent by N1 to N2, the synaptic behavior determines the way the information is treated. The synapse transforms a spike arriving from the presynaptic neuron into a chemical discharge of neurotransmitters (NT) detected by the post-synaptic neuron and transformed into a new spike. Markram and Tsodyks[24, 25] have proposed a phenomenological model to describe the synapse behavior. The synapse possesses a finite amount, $U$, of resources: the chemical neurotransmitters. Each spike activates a fraction $aU$ (a<1) of these resources and the amplitude $I$ of the transmitted spike is a function of this fraction. The fraction of neurotransmitters spent to transmit the information is then recovered with a characteristic time $\tau_{rec}$ (typically in the range of a second). The response of a synapse to a train of pulses depends on the time interval between successive pulses that determines the amount of available NTs. Depending on the nature of the synapse, the response to a constant frequency train of pulses can be either depressing or facilitating (cf. Fig. 3-a). Moreover, the biological synapse can process dynamical information when the frequency of the train of pulses is changed (figure 3-b). In the case of a purely depressive synapse, a depressing behavior is obtained at a "high" frequency train of pulses (decrease of the NTs available due to the low recovery between each spike separated by a time interval < $\tau_{rec}$), while the response of the synapse increases (facilitating behavior) at a lower frequency (the NTs get enough time to recover completely). This property has been extensively studied in biological synapse and is referred to as Short Term Plasticity – STP.[20, 25] This simple behavior gives to the synapse the main property that is necessary[2] for dynamical processing of information.

Varela et al. have developed a simple iterative model[26] to simulate the STP of biological synapses. Based on the work of Magleby et al.,[27] they describe the amplitude $I$ of a given spike in the post-synapse neuron by

$$I = \tilde{I}F_1...F_n D_1...D_m \qquad (1)$$





where the $F$ and $D$ terms are attributed to different facilitating or depressing mechanisms with specific time constants. Thus, the output of the simplest depressing synapse follows $I = \tilde{I}D$, where only one depressive term is considered (this expression is also used by Abbott et al. in [20]). $\tilde{I}$ is the intrinsic spike intensity delivered by the synapse (i.e. the intensity of the first spike after a long period of rest of the synapse) and $D$ is a dynamical factor representing the depression ($0 < D < 1$). The response of a synapse to a train of pulses at variable frequency is calculated by an iterative model: at each pulse arriving to the input, the output is calculated by combining the depression factor $D$ and the amount of NTs recovered (with a time constant $\tau_{rec}$) between two successive pulses. This quantitative model has shown a good agreement with the biological synapse behavior. We present in figure 3-b the comparison between the biological synapse response and the iterative model of Varela et al. with one facilitating and two depressing contribution (from [26]). We now demonstrate how the NOMFET can well mimic these typical synapse behaviors.

## 4. NOMFET - device behavior results and synapse analogy.

We first express the NOMFET output current taking into account the effect of the charged NPs. The pentacene being a p-type semiconductor, the NOMFET is active for negative gate voltage only. Such voltage has two effects. (i) As for usual transistors, it creates holes in the pentacene thin film at the interface with the insulator. (ii) It charges positively the gold particles via the organic material. Adapting the percolation theory of Vissenberg and Matters[28] for thin film organic transistors, the drain/source current of the NOMFET is expressed as (see supporting information)

$$I_{DS} = GV_{DS} \quad with \quad G = A_0 e^{\beta \varepsilon_F} e^{-\beta \Delta} \tag{2}$$

where $G$ is the channel conductance, $\beta = 1/k_B\Theta$, $\Theta$ the temperature, $A_0$ is a temperature dependent parameter, $\varepsilon_F$ is the Fermi energy fixed by the gate voltage and $\Delta$ is the shift of the Fermi energy induced by the charged NPs. The last term is caused by the repulsive





electrostatic interaction between the holes trapped in the NPs and the ones in the pentacene. The effect of the positively charged NPs is to reduce the current. To do an analogy with the biological synapses, the holes play the role of the neurotransmitters. They modify the output signal, $I_{DS}$, in a way that depends on the numbers of trapped charges via the $\Delta$ term: the more charges are stored in the NPs the more the current is reduced. Note that with electrons trapped in the NPs, the result is the opposite: the charges then increase the Fermi energy and the current. These behaviors correspond to our experimental observations reported elsewhere.[18] Charging NPs with holes by applying a negative voltage pulse on the gate induces a negative threshold voltage shift as measured on the drain current – gate voltage ($I_D$-$V_G$) curves of the NOMFET, and (at fixed $V_D$ and $V_G$), a decrease in the current. A positive gate voltage pulse shifts the NOMFET characteristics backward.[18] From these threshold voltage shifts (table 1) and for the optimized NOMFET with a NP density of around $10^{11}$ cm$^{-2}$ (such as those shown in Fig. 1-b to d), we estimated the number of charge per NP (for a gate voltage pulse of -50 V during 30 s, measured on a 5 µm channel length NOMFET at $V_D$ = - 30 V). They are typically ~2, ~6 and ~15 holes/NP for the 5nm, 10nm and 20 nm NP NOMFET, respectively.

Using these properties, we can mimic the different behaviors of the biological synapses by initially charging the NPs with holes (negative gate voltage) or discharging the NPs (positive gate voltage) depending whether we want a depressing or a facilitating behavior, respectively, before measuring the response of the NOMFET to a train of spikes. The input signal is the drain-source voltage, the output is the drain/source current (the transistor is used in a pass-transistor configuration, source and drain are reversible). The gate has two functions. In a first phase (see a chronogram of the signals in the supporting information, fig. S2), we used it to program the NOMFET by applying a gate voltage pulse $V_P$ while source and drain are grounded, and during the working phase, we applied a dc voltage $V_G$ to maintain the transistor in its "on" regime while an input spiking voltage (between $V_{D1}$ and $V_{D2}$) is applied





on the drain. Figures 4-a and b demonstrate that the facilitating and depressing behaviors are obtained on the same NOMFET, for rigorously the same application of pulses at the input, by initially programming the device with the gate signal $V_P$. For the depressing case (Fig. 4-a), we applied $V_P > 0$ before to run the device, and $V_P < 0$ for the facilitating case (Fig. 4-b). In a biological synapse, the facilitating behavior means that an incoming signal with a given frequency and duty cycle induces a post-synaptic signal having an increasing trend, whereas in the case of a depressing synapse, the post-synaptic signal tends to decrease as shown in Fig. 3-a. We demonstrated in figure 4 exactly the same behavior for the NOMFET, where the programming negative/positive gate voltage pulse induces the facilitating/depressing behavior, respectively. Interestingly, the same behavior (Fig. 5) is also obtained for NOMFET with a less controlled deposition of NPs (i.e. NPs forming aggregates, Fig. S3 in supporting information). This result means that the synaptic behavior of the NOMFET is robust against process variations.

More importantly, we can reproduce the STP behavior without initial programming. We used the NOMFET as a "pseudo two-terminal device". The gate receives the same input voltage (a train of pulses at frequency $1/T$, $T$ is the period, and amplitude $V_P$) as the drain/source electrode (supporting information, Fig. S2). We measured the response of the NOMFET to sequences of pulses with different periods, $T$ (Fig. 6). During such experiments, the NPs are alternatively charged during the pulse time and discharged between pulses.[18] The value of the current at a certain time (i.e. after a certain number of spikes) depends on the past history of the device that determines the amount of charges presents in the NPs. To illustrate this point, let us consider the NOMFET (L = 12 μm, 5 nm NPs) at the beginning of a particular sequence with period T (Fig. 6-a), where the NPs contain some charges. If T $<<$ $\tau_d$ ($\tau_d$ is the NP discharge time constant of about 20 s here, see below), more and more holes are trapped in NPs and the NOMFET presents a depressing behavior as observed for the 0.5 and 2





Hz spike sequences. As expected, the depressing behavior is more pronounced when increasing the frequency. This result is also in good agreement with the behavior of a spiking biological synapse (see supporting information). Then, for a larger period T (0.05 Hz signal in Fig. 6-a), the NPs have enough time to be discharged between pulses and the sequence presents a facilitating behavior. This feature exactly reproduces the behavior of a biological synapse (Fig. 3-b). The holes trapped in the NPs play the role of the neurotransmitters and the output signal, $I_{DS}$, is a decreasing function of the number of holes stored in the NPs[18] - Fig. 6. At each spike, a certain amount of holes are trapped in the NPs. Between pulses the system relaxes: the holes escape with a characteristic time $\tau_d$. We did not observe such a synaptic behavior for the reference pentacene OFET (no NPs, see supporting information, Fig. S4).

Based on eq. (2) for the output current, we adapted the simplest iterative model of Varela et al.[26] to simulate the dynamical behavior of the NOMFET, and we developed an analogous modelisation . A pulse induces positive charges in the NPs: the NOMFET channel conductance is then reduces by a multiplicative factor K < 1. Between each pulses, the NPs tends to discharge with a characteristic time $\tau_d$. The general equation describing the iterative model is (see supporting information):

$$I_{n+1} = I_n K e^{-(T-P)/\tau_d} + \tilde{I}\left(1 - e^{-(T-P)/\tau_d}\right)K \qquad (3)$$

where $I_{n+1}$ and $I_n$ are the current of the NOMFET at the end of the n+1[th] and n[th] voltage pulses sent to the NOMFET, $\tilde{I}$ is the intrinsic drain current of the NOMFET i.e. the current that would have the device if the charges in the NPs would be kept in the equilibrium configuration (i.e. after a long period of rest, with no charge, or with a constant residual charge induced by the static dc bias of the device), $T$ is the period between two pulses and $P$ is the width of the pulses.

We fitted the iterative model to simulate the NOMFET behavior (red circle on Fig. 6). The same fitted $\tilde{I}$, $K$ parameters and $\tau_d$ (here 4.1x10[-9] A, 0.9 and 20 s) are used in the three





successive sequences shown in Fig. 6-a proving a good agreement between the model and the experiments. For instance, let us consider the system at the beginning of a particular sequence with period $T$ (T =0 .5 s or F = 2 Hz in Fig. 6-a), where the NPs contain some charges. Since $T << \tau_d$, the first term of the iterative function, $I_n K e^{-(T-P)/\tau_d}$ , is the more important one and the sequence of spikes present a depressing behavior ($I_{n+1} < I_n$). At the opposite, for a larger period $T$ (T = 20 s, F = 0.05 Hz), the second term $\tilde{I}\left(1 - e^{-(T-P)/\tau_d}\right)K$ becomes the larger one and the sequence presents a facilitating behavior ($I_{n+1} > I_n$). Again, the depressing behavior for T = 2 s (F = 0.5 Hz) is well reproduced by the model. We applied this model to different NOMFET with various W/L ratio (L = distance between electrodes and W = width of the electrodes) down to L = 200 nm. For instance, the fits shown in Fig. 6-b for the L = 2μm NOMEFT correspond to $\tau_d = 3\ s$ ( $\tilde{I} = 10^{-6}$ A, $K = 0.99$) and for a L = 200nm NOMFET (Fig. 6-c) to $\tau_d = 0.9\ s$ ( $\tilde{I} = 6.3\text{x}10^{-10}$ A, $K = 0.98$). We report in Fig. 7 the evolution of $\tau_d$ with the channel length and size of the NPs. For comparison, we also plot (for some devices) the time constant extracted from basic charge/discharge measurements (see supporting information, Fig. S5). We note that: (i) the characteristic time from the fitted model and the direct discharge measurement are of the same order of magnitude, (ii) the characteristic time, thus the working frequency range of the NOMFET, can be adjusted by changing the channel length L of the NOMFET, (iii) $\tau_d$ is weakly dependent of the NPs size. The RC charge/discharge time constant is roughly governed by the channel resistance of the NOMFET, which scales as L, and the self-capacitance of a NP $C_{self} = 2\pi\varepsilon D,$ which scale as the NP diameter D (ε the dielectric constant), thus scaling down L and D should, in principle, decrease $\tau_d$. This feature is clearly observed for L. At the macroscopic scale of the NOMFET, we have to take into account the total capacitance, which is $NC_{self}$ with N the number of NPs in the channel. This number, while controlled to be around $10^{11}$ NP/cm$^2$ (see Fig. 1), varies from device to device, and we believe that these dispersions can hinder the intrinsic role of the





NP size. Moreover, the NPs are capped with surfactant molecules (see experimental section), which can act as a tunnel barrier between the NPs and the pentacene. It is likely that this tunnel barrier play a role in the discharge phenomena (a thicker tunnel barrier will increase discharge time constant). We did not control the structural quality of this tunnel barrier, which can represent an additional source of dispersion. More detailed experiments, for instance by systematically varying the length of the alkanethiol capping the NPs, will be necessary to increase the control of the charging/discharging phenomena in the NOMFET. A more accurate description of the relationship between these time constants, the size of the NPs and geometry of the NOMFET would require a more sophisticated modelization using a 2D network of distributed RC time constants (taken into account statistical dispersion) rather than a single macroscopic one.

Finally, we note that the iterative model used here to fit the experimental data can be implemented in usual device simulator (SPICE-like) allowing a reliable conception and simulation of hybrid NOMFET/CMOS neuronal circuits. [29] These device/circuit simulations can easily take into account the experimental dispersion of the NOMFET performances to test the robustness of these neuronal circuits against the actual device variation. Recently, a simple associative memory has been built using a purely CMOS-based emulator of memristor (acting as the synapse) and neurone.[30] In this work, the memristor-synapse is emulated using a combination of CMOS analog-to-digital converter and microcontroller. It is likely that the NOMFET can be used as the synapse in such associative memory architecture. Such hybrid NOMFET/CMOS neuromorphic computing circuits and architectures are currently under investigation in our laboratory. At this stage, an interesting question can be raised, whether the NOMFET is more related to a memristor device[9] or more related to a memcapacitor as defined by Di Ventra et al. [31] While a definitive answer would probably require more experiments and simulations of the NOMFET, we believe that the NOMFET is more related





to a charge-controlled memristor (or charge-controlled memconductance) as defined in Ref.

[31], since the channel conductance of the NOMFET is history-dependent of the amount of

charges stored in the NPs.

**5. Conclusion.**

In conclusion, we demonstrated a hybrid nanoparticle-organic device, a NOMFET, that makes

use of the charge storage capability of the nanoparticles and the amplification factor of the

organic transistor to mimic the short-term plasticity of a biological synapse. The NOMFET

can be programmed to exhibit both a facilitating or depressing behavior. By adjusting the size

of the device down to 200 nm and the diameter of the nanoparticle down to 5 nm, we can

range the working frequency between 0.01 Hz and 10 Hz. We simulated the synapse behavior

of the NOMFET adapting a model developed for the biological synapse. Varela et al.[26]

describe the output of biological synapses as a product of several depression (*D*-terms) and

facilitation (*F*-terms) factors (eq. 1), each of these factors being associated with a particular

characteristic time. For the NOMFET, by approximating the relaxation function of the NP

discharge by a simple exponential, we get the simplest model of this type: our NOMFET

behaves within this approximation as the simplest depressive synapse with only one *D*-term.

**6. Experimental**

*Fabrication of devices*: The NOMFETs were processed using a bottom-gate electrode

configuration. We used highly-doped (~$10^{-3}$ Ω.cm) p-type silicon covered with a thermally

grown 200 nm thick silicon dioxide. Before use, these wafers were cleaned by sonication in

chloroform for 5 min then by a piranha solution ($H_2SO_4$/$H_2O_2$, 2/1 v/v) for 15 minutes and

then by ultraviolet irradiation in an ozone atmosphere (ozonolysis) for 30 minutes (***Caution:***

***preparation of the piranha solution is highly exothermic and reacts violently with organics***).

Electrodes (titanium/gold (20/200 nm)) were deposited on the substrate by vacuum





evaporation and patterned by optical lithography (for channel length L between 1 and 20 µm). Smaller NOMFETs (L = 0.2 µm) were fabricated on thinner oxide (10 nm thick) by usual electron-beam lithography.

Then, the $SiO_2$ (gate dielectric) was functionalized by self-assembled monolayer (SAM) to anchor gold nanoparticles (NPs) on the surface. For the largest NPs (20 and 10 nm in diameter), the $SiO_2$ surface was functionalized by an amino-terminated SAM before the NP deposition.[13, 32, 33] First, gold (Au) electrodes were functionalized by dipping in a 2-amino ethanethiol solution in ethanol (10mg/mL) during 5h. The sample was then rinsed 3 times with isopropanol and subsequently dried in argon stream. Second, the $SiO_2$ surface was functionalized by immersion in a solution of (3-aminopropyl) trimethoxysilane (APTMS) molecules (supplied by ABCR) diluted in anhydrous toluene at a concentration 1.25µL/mL and at 60°C during 4min.[34] The reaction took place in a glove-box with a controlled atmosphere (nitrogen, with less than 1 ppm of oxygen and water vapor). Excess, non-reacted, molecules were removed by rinse in toluene, and then in isopropanol under sonication. This sample was subsequently dried under argon stream. Static water contact angle was 19°, a common value for hydrophilic $NH_2$-terminated surfaces.[34] This sample was then dipped in an aqueous solution of citrate-stabilized Au-NP (colloidal solution purchased from Sigma Aldrich, 20 ± 3 nm and 10 ± 3 nm in diameter) overnight under argon atmosphere. NP concentration in the solution and duration of the reaction are changed to adjust the NP density on the surface. The sample was then cleaned with deionized water and isopropanol, and dried under argon stream. Finally, Au-NP were encapsulated by dipping in a solution of 1,8-octanedithiol (from Aldrich) in ethanol (10µL/mL) during 5h to help the formation of a network of NPs. The sample was subsequently rinsed in alcohol and dried in argon stream. For the smallest NPs, we used a solution of 4-5 nm (in diameter) dodecanethiol functionalized gold nanoparticles (2 % in toluene) supplied by Aldrich. The oxidized silicon with gold





electrodes wafer was placed overnight in the presence of vapors of freshly distilled

mercaptopropyltrimethoxysilane (MPTS) in a laboratory glassware at 0.2 Torr. [35, 36] This

freshly prepared substrate was immersed in a gold nanoparticles (NPs) solution. The starting

solution supplied by Aldrich was diluted 100 times in toluene. Again, NP concentration in the

solution and duration of the reaction are changed to adjust the NP density on the surface. As

expected, thiol capped Au NPs readily react with thiol-terminated SAM by ligand exchange

forming a covalent bond with the surface.

Finally, a 35 nm thick pentacene film was evaporated at a rate of 0.1 Å/s. The

substrate was kept at room temperature. A reference device of pentacene without NPs (and

without SAM) was also realized in the same run of deposition to evidence the effect of NPs

on the electrical properties. Note that a second reference sample with a SAM between the

$SiO_2$ and the pentacene film did not show any memory effect. [37]

*Electrical measurements*: The NOMFET electrical characteristics were measured with an

Agilent 4155C semiconductor parameter analyzer, the input pulses were delivered by a pulse

generator (Tabor 5061). The electrodes of the NOMFET were contacted with a micro-

manipulator probe station (Suss Microtec PM-5) placed inside a glove box (MBRAUN) with

a strictly controlled nitrogen ambient (less than 1 ppm of water vapor and oxygen). Such a dry

and clean atmosphere is required to avoid any degradation of the organics (SAMs and

pentacene). So the devices are not exposed to air from the beginning of the fabrication process

to the end of the electrical characterization.

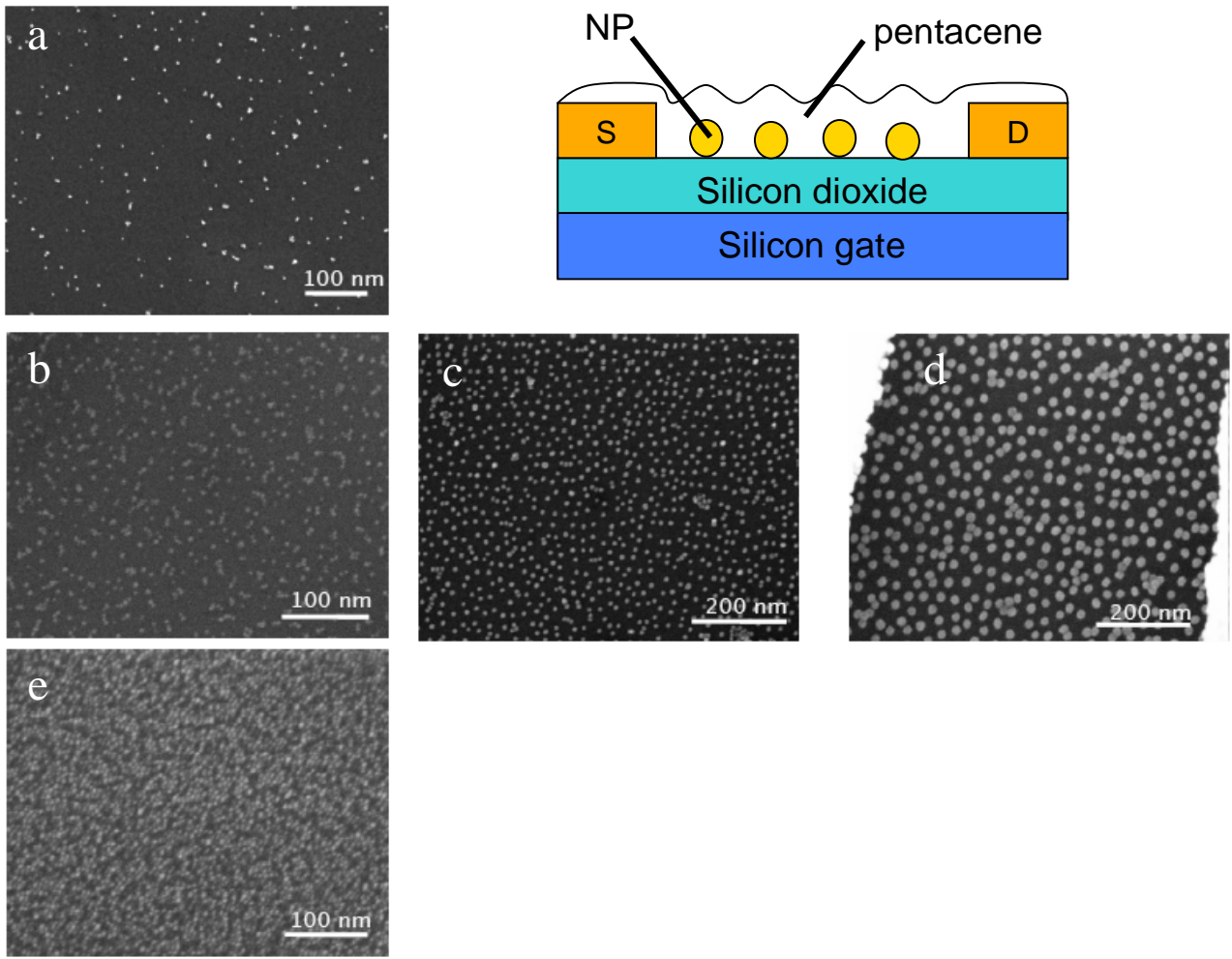

**Figure 1**. Scanning electron microscope image of the NP arrays before the pentacene deposition. **a)** 5 nm NP with a low density of 4.4x10^{10} NP/cm², **b-d)** 5, 10 and 20 nm NP, respectively, with a medium density of 3.7x10^{11}, 1.8x10^{11} and 0.9x10^{11} NP/cm², respectively, **e)** 5 nm NP with a high density >10^{12} NP/cm² showing a quasi 2D film with a strong percolation.





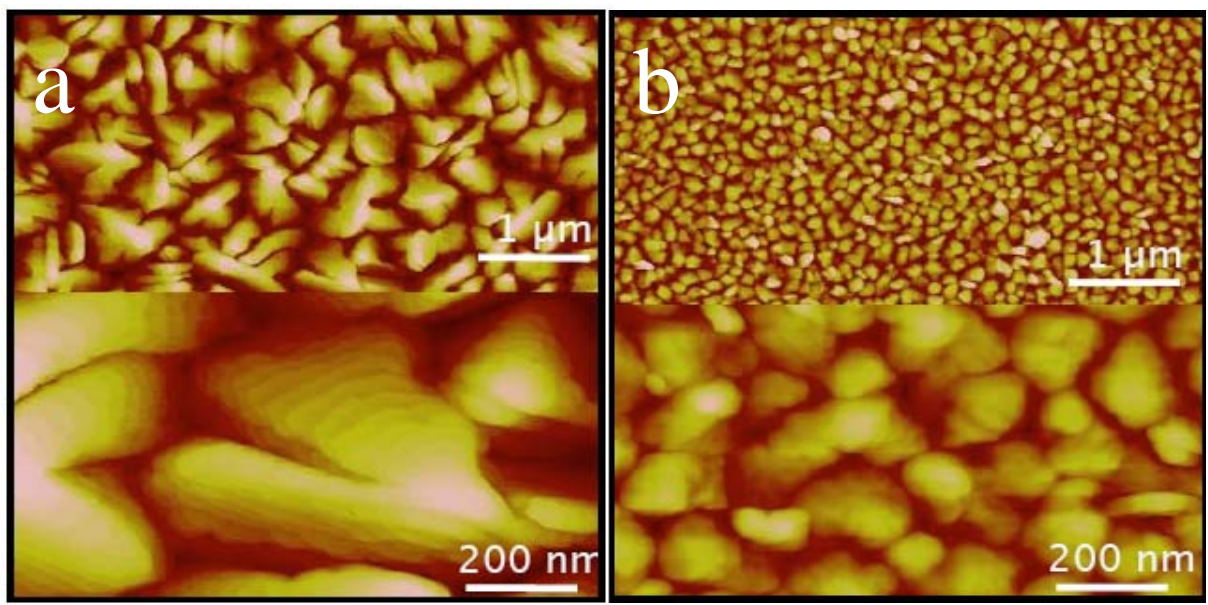

**Figure 2**. TM-AFM images of the 35 nm thick pentacene films at two magnifications. **a)** without the NPs and **b)** with an array of 5 nm NPs at a density of around $10^{11}$ NP/cm$^2$ as in figure 1-b.





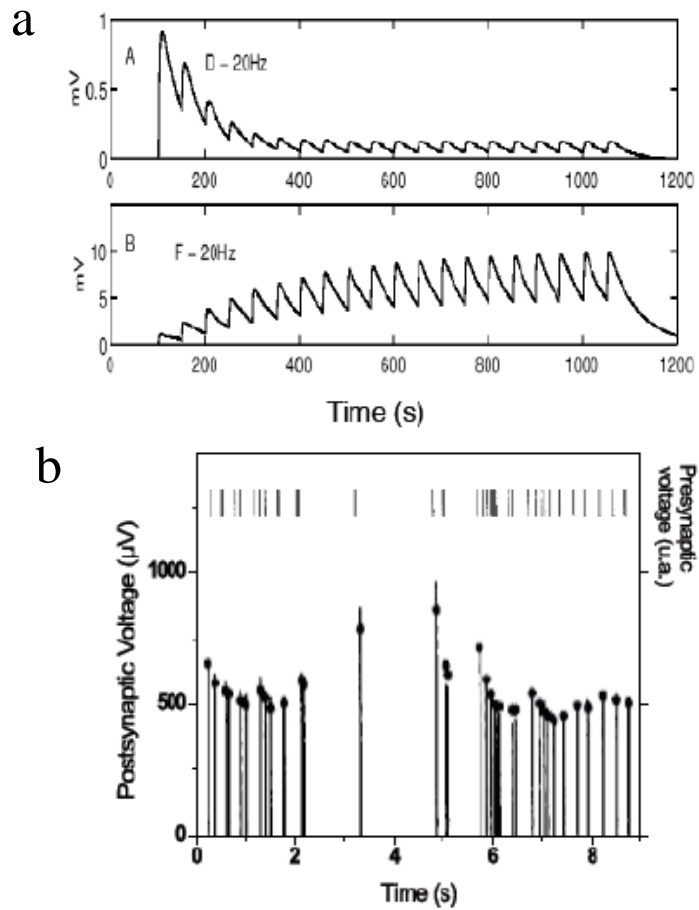

**Figure 3. a)** simulation of post-synaptic potential for a depressing (top figure) and a facilitating (down figure) synapse submitted to the same pre-synaptic pulse train at 20 Hz (adapted from Ref. [24]). **b)**, comparison between the frequency-dependent post-synaptic potential response of a depressing synapse (lines) and the iterative model of Varela et al. (dots), adapted from Ref. [26] as a function of frequency of the pre-synaptic input signal.





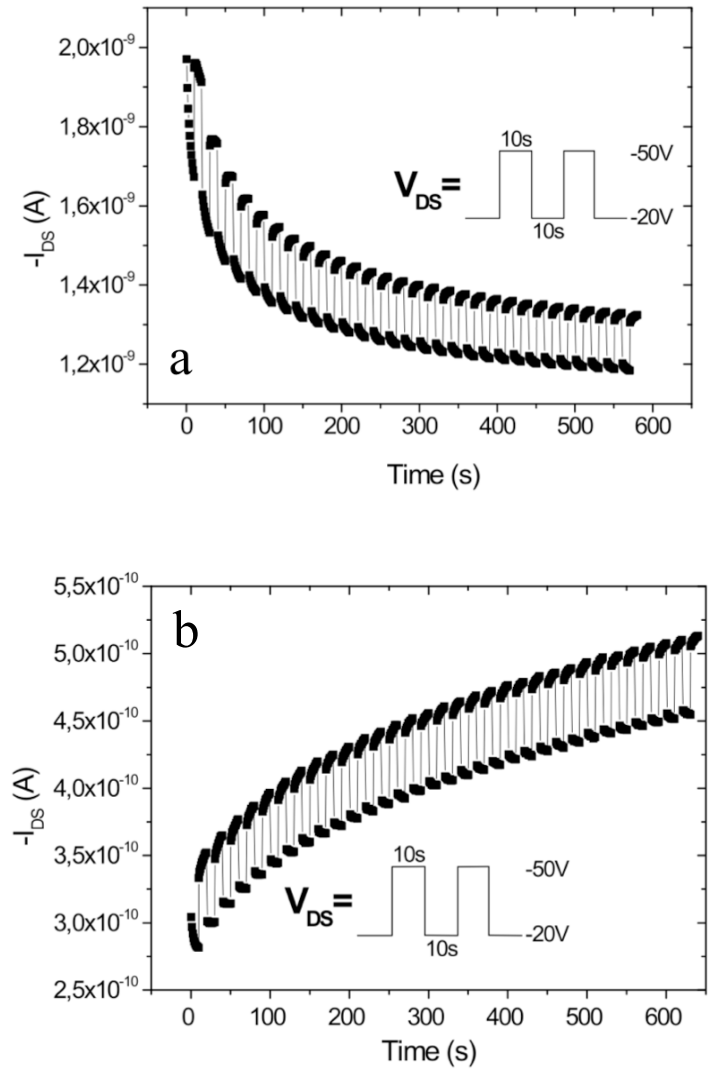

**Figure 4. a**) response (drain current) of the NOMFET (5 nm NP, L/W = 12 μm/113 μm) to a constant frequency (0.05 Hz), train of pulses ($V_{D1}$ = -20 V, $V_{D2}$ = - 50V) after programming the device by a $V_P$ = 50V (40s) gate pulse to discharge the NPs. The decreasing trend of the output current (in absolute value) mimics a depressing biological synapse. **b**) response (drain current) of the NOMFET (5 nm NP, L/W = 12 μm/113 μm) to a constant frequency (0.05 Hz) train of pulses ($V_{D1}$ = -20 V, $V_{D2}$ = - 50V) after programming the device by a $V_P$ = -50V (40s) gate pulse to charge the NPs. The increasing trend of the output current (in absolute value) mimics a facilitating biological synapse.





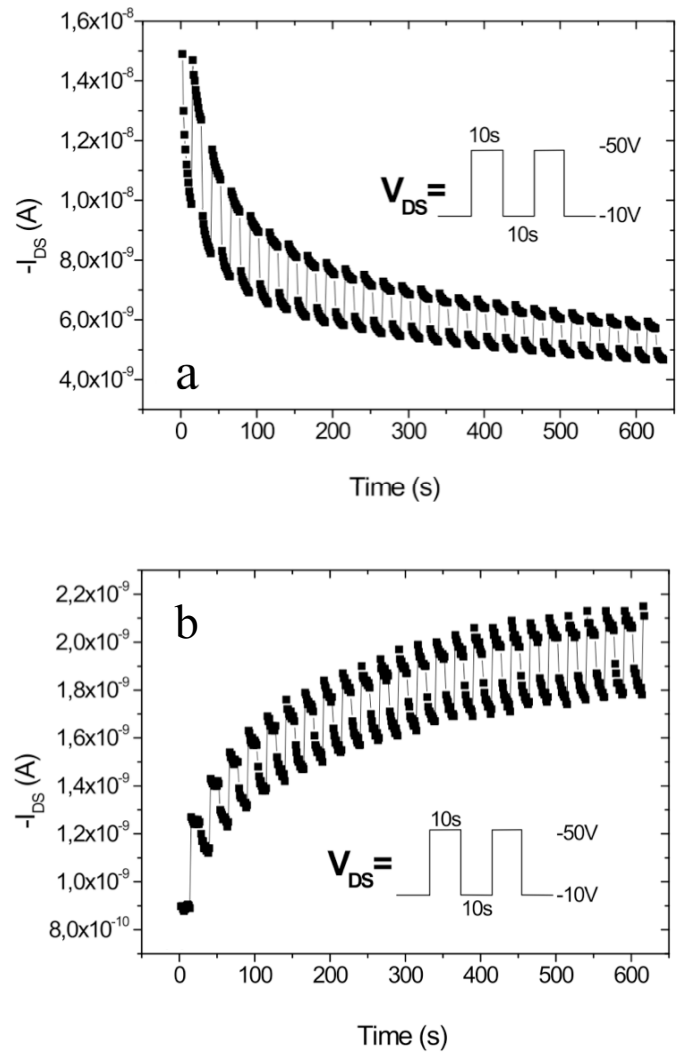

**Figure 5**. **a)** response (drain current) of the NOMFET (20 nm NP, L/W = 12 μm/113 μm) to a constant frequency (0.05 Hz), train of pulses ($V_{D1}$ = -10 V, $V_{D2}$ = - 50V) after programming the device by a $V_P$ = 50V (20s) gate pulse to discharge the NPs. The decreasing trend of the output current (in absolute value) mimics a depressing biological synapse. **b)** response (drain current) of the NOMFET (20 nm NP, L/W = 12 μm/113 μm) to a constant frequency (0.05 Hz) train of pulses ($V_{D1}$ = -10 V, $V_{D2}$ = - 50V) after programming the device by a $V_P$ = -50V (20s) gate pulse to charge the NPs. The increasing trend of the output current (in absolute value) mimics a facilitating biological synapse.





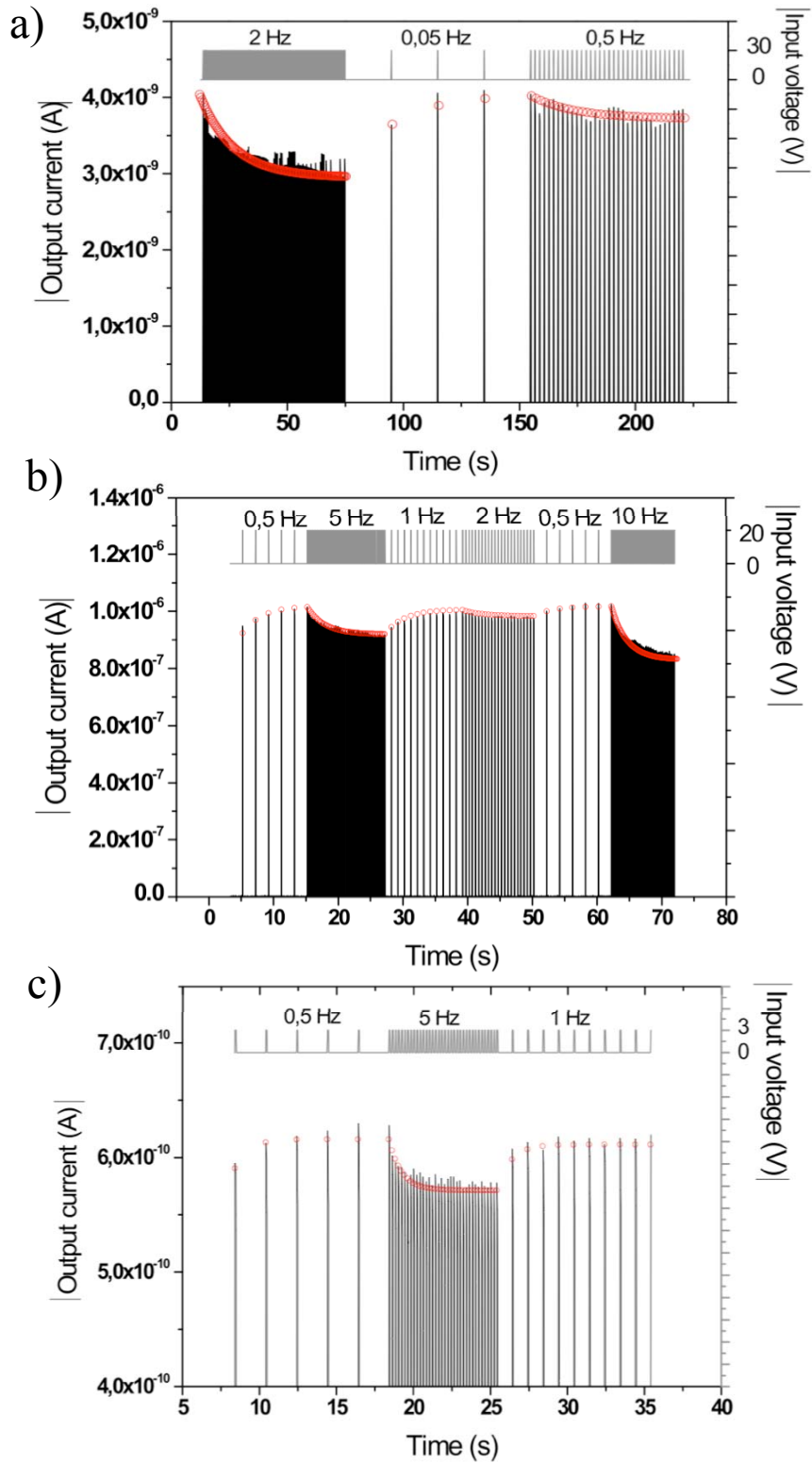

**Figure 6. a)** response (drain current) of NOMFET with L/W ratio of 12 µm/113 µm and NPs size of 5 nm to sequences of spikes at different frequencies (pulse voltage $V_P$ = - 30V). **b)** response of NOMFET with L/W ratio of 2µm/1000µm and NPs size of 5 nm (pulse voltage $V_P$ = - 20V). **c)** response of NOMFET with L/W ratio of 200nm/1000µm and NPs size of 5 nm (pulse voltage $V_P$ = - 3V). In both cases, the red circles correspond to the iterative model calculation (see text), the black lines are the output current measurements. Increasing the





frequency evidences a depressing behavior and a facilitating one is observed by decreasing the frequency.

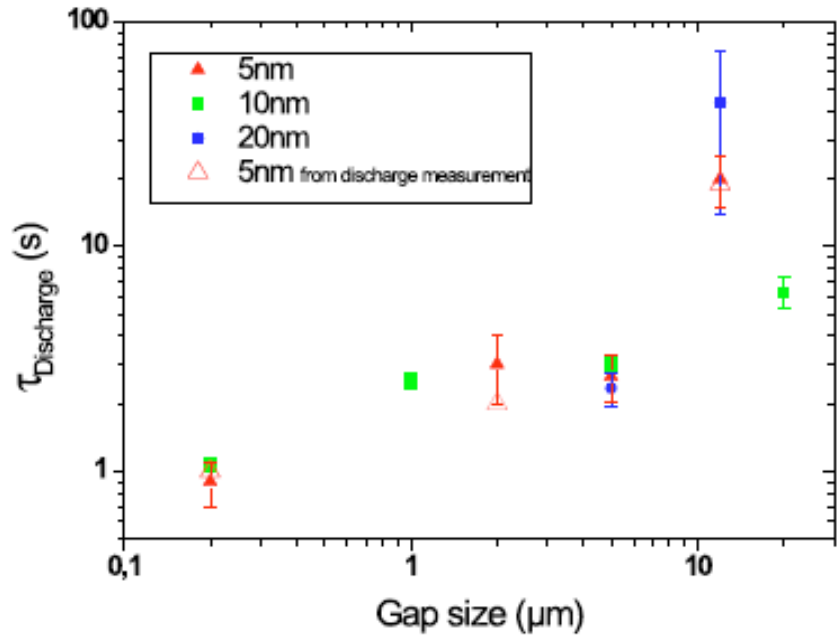

**Figure 7**. Evolution of the NP discharge time constant as function of the NOMFET channel length and for the different sizes of NPs. Full symbols are from data extracted by fitting the model on the experiments (as in Fig. 6), open symbols are from discharge experiments (supporting information, Fig. S5).





|  | Reference | 5 nm NP | 10 nm NP | 20 nm NP |
|---|---|---|---|---|
| Average Id (A) | $1.8x10^{-4}$ ($\pm 8.5x10^{-5}$) | $2.3x10^{-6}$ ($\pm 1.6x10^{-5}$) | $4.1x10^{-7}$ ($\pm 7.5x10^{-7}$) | $3.3x10^{-7}$ ($\pm 2.8x10^{-7}$) |
| Average mobility (cm$^2$V$^{-1}$s$^{-1}$) | 0.13 ($\pm 0.05$) | $2x10^{-3}$ ($\pm 2x10^{-3}$) | $7.6x10^{-4}$ ($\pm 7.9x10^{-4}$) | $2.0x10^{-3}$ ($\pm 2.9x10^{-3}$) |
| Vt shift (V) | 1.1 ($\pm 0.4$) | 7.5 ($\pm 1.4$) | 9.4 ($\pm 0.9$) | 12.8 ($\pm 1.8$) |

**Table 1**. Main transistor parameters of the NOMFET with different NP size and of the reference device (no NP). All measurements for a L=5 µm NOMFET. Average values and error bars correspond to the mean and standard deviation of a normal distribution fitted on the experimental data. The threshold voltage shift is the difference between the threshold voltages after and before charging the NP by a negative pulse voltage applied on the gate (-50V during 30s). Note that the small value measured for the reference device is known and due to traps in the pentacene or at the SiO$_2$/pentacene interface (see Ref. 18 and reference therein).





**The table of contents entry.** Schematic representation between a biological synapse and the NOMFET (nanoparticle organic memory field effect transistor). Charges in the nanoparticles play the role of the neurotransmitters. The NOMFET reproduces the short-term plasticity of a spiking synapse and can be programmed to work as a facilitating or depressing synapse. The NOMFET is a nanocomputing building block with acquired behavior.



F. Alibart, S. Pleutin, D. Guérin, C. Novembre, S. Lenfant, K. Lmimouni, C. Gamrat and D. Vuillaume.

Title ((no stars))

ToC figure ((55 mm broad, 50 mm high, or 110 mm broad, 20 mm high))

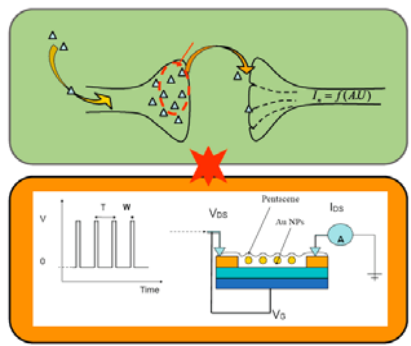

Column Title: *F. Alibart et al.*/Short title





**Supporting Information.**

# An organic-nanoparticle transistor

# behaving as a biological spiking synapse


F. Alibart,[1] S. Pleutin,[1] D. Guerin,[1] C. Novembre,[2] S. Lenfant,[1] K. Lmimouni,[1]

C. Gamrat,[2] & D. Vuillaume.[1*]

1. Molecular Nanostructures and Devices group, Institute for Electronics Microelectronics and

Nanotechnology, CNRS, University of Lille, BP60069, avenue Poincaré, F-59652cedex, Villeneuve

d'Ascq, France.

2. CEA, LIST/LCE (Advanced Computer technologies and Architectures), Bat. 528, F-91191, Gif-sur-

Yvette, France.

* dominique.vuillaume@iemn.univ-lille1.fr


## *Supporting information*

## I. Drain current of the NOMFET, role of the NPs

The transport properties of organic thin film transistors are reasonably well understood in terms of a percolation theory developed already ten years ago by Vissenberg and Matters (VH)[1]. The advantage of this theory is to provide an analytical expression for the conductance of the transistor channel. We start by summarizing the main features of the model. A thin film of pentacene is not perfectly organized, and it is likely that the disorder increases in the presence of the network of NPs. For small bias (linear regime) the organic film can be viewed as an electrical network with a quantum localized state of energy $\varepsilon_i$ at each node i.[2] The site energies are exponentially distributed. The density of states (DOS) is given by

$$g(\varepsilon) = \frac{N_t}{k_B \theta_0} \exp\left(\frac{\varepsilon}{k_B \theta_0}\right) \tag{S1}$$





with $N_t$ the number of localized state per unit volume, $k_B\theta_0$, a measure of the energy width of the distribution (Fig. S1). The energies are such that $-\infty < \varepsilon \leq 0$. It is important to note that this type of DOS introduced by VH is not arbitrary and has been recently observed by Kelvin probe force microscopy for thin films of copper phthalocyanine[3], for instance.

Each pair of nodes, i and j, distant by $R_{ij}$ is connected by a bond with a conductance $G_{ij}=G_0 exp(-S_{ij})$ where[1,2]

$$S_{ij} = 2\alpha R_{ij} + \frac{\left|\varepsilon_i - \varepsilon_F\right| + \left|\varepsilon_j - \varepsilon_F\right| + \left|\varepsilon_i - \varepsilon_j\right|}{2k_B\theta} \tag{S2}$$

The first term takes account for usual tunneling processes and the second for thermally assisted tunneling. $\alpha$ is an effective overlap parameter, $\varepsilon_F$ is the Fermi energy imposed by the reservoirs (electrodes) which fixes the number of careers and $\theta$ is the temperature. The second term is crucial to understand the conductivity in organic thin film transistor. It selects the quantum states with energy sufficiently close to the Fermi level to participate to the charge transport: the more the DOS is high at the Fermi level, the more the conductance will be high. Because of the exponential DOS, by increasing the gate voltage we increase the number of states at the Fermi level and consequently the conductance. This is the usual behavior for an organic transistor but, in this case, the conducting channel is disordered (more or less random distribution of $R_{ij}$). Solving the percolation problem, VH have found the channel conductance

$$G = G_0\left(\frac{\pi N_t(\theta_0 / \theta)^3}{(2\alpha)^3 B_C}\right)e^{\beta\varepsilon_F} = A_0 e^{\beta\varepsilon_F} \tag{S3}$$

where $G_0$ is a parameter and $B_C$ a constant (~2).

In the NOMFET, when the gold particles are positively charged, there is a Coulomb repulsion between the trapped charges in the NPs and the holes created in the pentacene thin film by the gate voltage. We assume that the only effect of the charged NPs is on the energy of the localized states. The gold NPs are assumed to be all perfectly identical. Every lattice site i feels a repulsive Coulomb interaction

$$\delta_i = \frac{q^2}{4\pi\varepsilon_0\varepsilon_r}\sum_\alpha \frac{n_\alpha}{\left|\vec{r}_\alpha - \vec{r}_i\right|} \tag{S4}$$





where the summation runs over all the NPs which are labelled by α. $n_\alpha$ is the number of holes stored in the NP α ($n_\alpha$=0, 1, 2,…), $q$ is the elementary electric charge, $\varepsilon_0$ and $\varepsilon_r$ denote as usual the permittivity of the vacuum and of the material, respectively. $\vec{r}_\alpha$ and $\vec{r}_i$ are the position vectors of the NP α and the quantum state i, respectively. Note that the summation does not show any divergence since the NPs and the quantum state never occupy the same spatial location. $\delta_i$ is a stochastic variable. We face a new percolation problem where all the site energies are shifted up by the repulsive terms, $\varepsilon_i \rightarrow \varepsilon_i + \delta_i$.

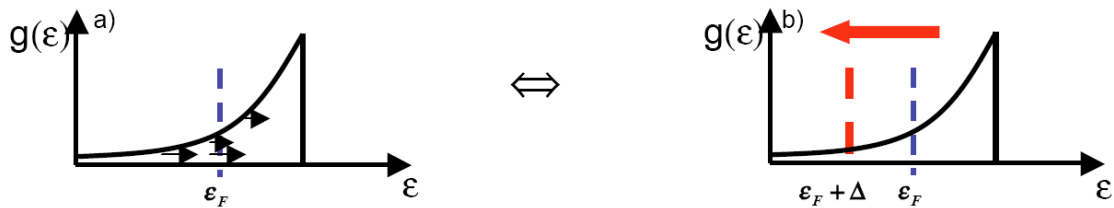

**Figure S1**. **a**, The quantum state energies are shifted up by the interaction with the charged NPs. **b**, Equivalently the Fermi level is shifted down by the same interaction.

When the NPs are positively charged the conductance of the pentacene thin film decreases. This is due to the exponential DOS. The Fermi level is fixed by the electrodes. The site energies are all shifted up by the repulsive Coulomb interaction (Figure S1) and consequently the number of states at the Fermi level is reduced. Since we are only interested to calculate the conductance of the NOMFET this is, to a first approximation, the only important information. This decrease of the conductance can be simply modeled by keeping the unperturbed DOS and shifting down the Fermi level by a constant $\Delta$ (Figure S1). We further assume that the value of $\Delta$ depends only on the applied gate voltage. Choosing a particular voltage fixes the number of positive charges trapped in the NPs. It is clear however that the precise way these charges are distributed in the NPs should in principle influence the transport properties but, how this is done and how large is this dependence remain to be understood. Nevertheless, it turns out that our approximation is sufficient to model our data with a reasonable accuracy.

To summarize, the effect of the repulsive interaction due to positive charges in the NPs is to reduce the Fermi energy by an amount $\Delta$ that is related to the total number of charges trapped in the particles. In other terms, it is a function of $V_G$. Then





$$G = A_0 e^{\beta \epsilon_F} e^{-\beta \Delta} \qquad (S5)$$

meaning that the amplitude of the current, $I=GV_{sd}$, is reduced by the holes stored in the particles. This is indeed what is experimentally observed[4].

## II. Chronograms.

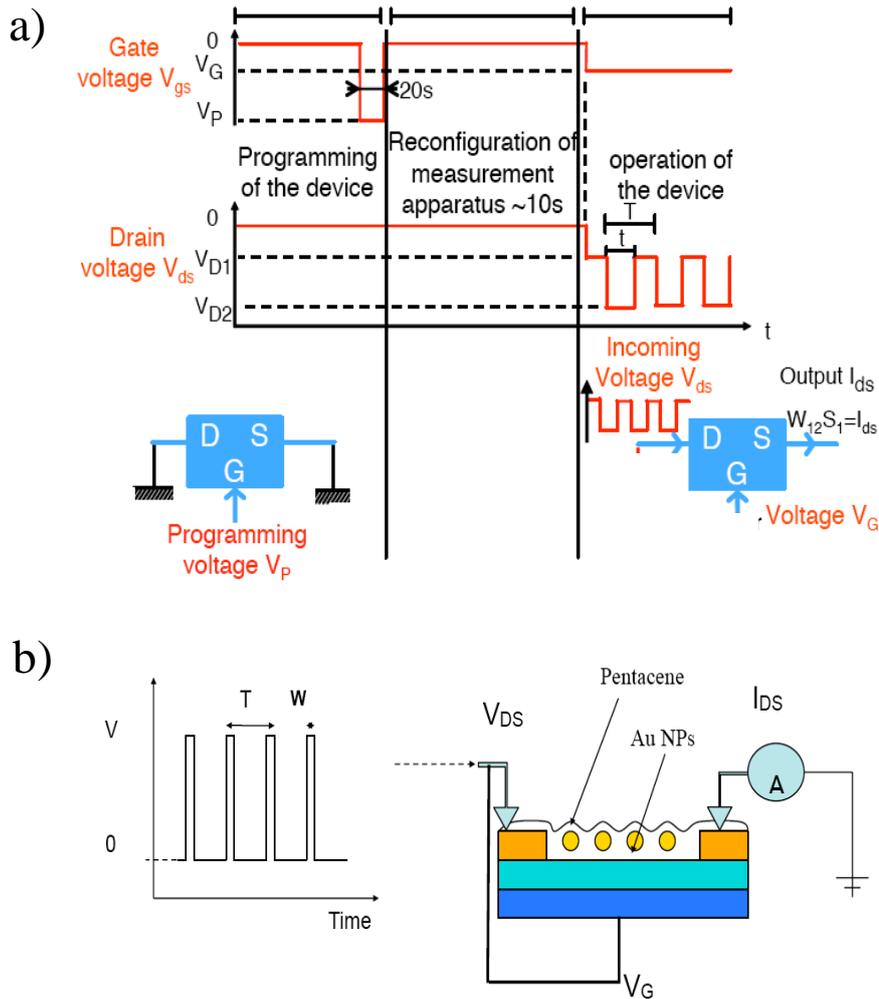

**Figure S2**. **a)** Typical gate and drain voltage chronogram for the programming period (pulse $V_P$ applied on the gate) and functioning period (a train of pulse voltages between $V_{D1}$ and $V_{D2}$ is applied on the drain/source, while the gate is dc biased at $V_G = -20V$). **b)** Typical signal chronogram for the STP experiment.





## III. Aggregated NP networks.

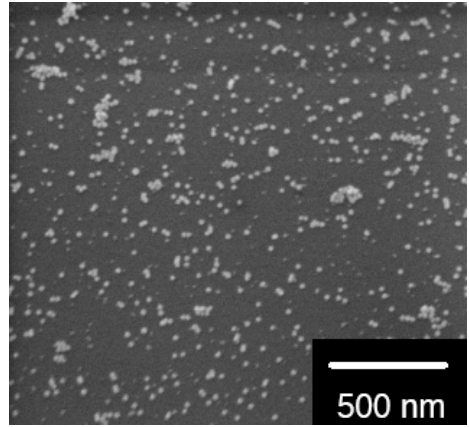

**Figure S3.** Scanning electron microscope images of aggregated 20 nm NP networks such as those used in the NOMFET reported in Fig. 5.

## IV. Reference device (no NP)

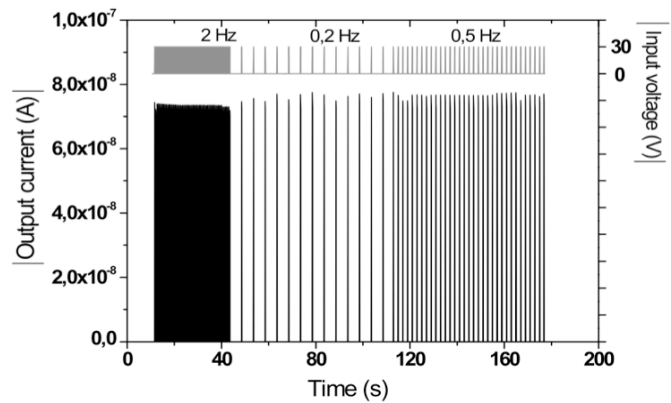

**Figure S4.** Same experiment as in Figure 6 (main text) for a reference device without NPs and W/L ratio of 12µm/113µm. We did not observe a STP behavior.

## V. Iterative model adapted to the NOMFET

The next step is to model the response of the NOMFET to a sequence of spikes. Here, we first consider a train of pulse voltage of frequency *1/T* and amplitude $V_P$, each pulse has a time width $P$ (as in Figs. 4-6 in the main text). To model the NOMFET response, we consider the iteration $I_n$, the value of the current at the end of the $n^{th}$ pulse. During the pulse the particles are charged with a certain





dynamic characterized by a function $f_c(t)$, t being the time variable. For a pulse with an infinite width

the charges stored in the NPs is such that we get the exponential prefactor $e^{-\beta\Delta}$ in the conductance.

We further assume to be as close as possible to Varela's model[5] developed for biological synapses

(see main text), that the amount of charges entering the NPs at each spike depends only on the

properties of the pulse ($V_P$, the amplitude, $P$, the width) but not on the history of the device.

Overcoming this constraint does not change significantly the results for the typical sequences of

spikes considered in this work: the approximation of constant charging gives very good agreements

with experiments (Fig. 6, main text). Between pulses, the particles are discharged according to a

dynamic described by a function $f_d(t)$. For an infinite period, the particles are completely empty giving a

factor 1 in the conductance. Collecting these informations together we get the general iteration

$$I_{n+1} = I_n\left(e^{-\beta\Delta} - \left(e^{-\beta\Delta} - 1\right)f_c(P)\right)f_d(T-P) + \tilde{I}\left(1 - f_d(T-P)\right)\left(e^{-\beta\Delta} - \left(e^{-\beta\Delta} - 1\right)f_c(P)\right)$$ (S6)

where $\tilde{I}$ is the current that would carry the device within the same voltage conditions but without NPs.

The two relaxation functions are such that $f_c(0) = f_d(0) = 0$. The starting value $I_0$ of the iteration depends

on the initial charging condition of the NPs (that can be adjusted by an initial pulse on the gate). Since

we consider only pulses with constant width $P$ and amplitude $V_P$ during a sequence (only the period

varies), we can replace the charging part of the iteration by a constant multiplicative factor

$$K = e^{-\beta\Delta} - (e^{-\beta\Delta} - 1)f_c(P)$$ (S7)

The discharging function $f_d(t)$ of the NPs is then evaluated by fitting the iterative model to the

sequences of pulses measurements such as those shown in Fig. 6 (main text). This function is

approximated by a single exponential with a reasonable accuracy

$$f_d(t) \approx f_d(0)e^{-t/\tau_d}$$ (S8)

The resulting iterative function that will be used for the modeling has the simple form:

$$I_{n+1} = I_n K e^{-(T-P)/\tau_d} + \tilde{I}\left(1 - e^{-(T-P)/\tau_d}\right)K$$ (S9)

It is important to note that with this simplification, good agreements are obtained with experiments

even if the discharging function should be more complex, and we are in the same approximation limit

than Abbott et al.[6] when these authors discussed the role of short term plasticity in biological synapses.

As a matter of illustration, let us write down the two very first iterations for the case where

$I_0 = \tilde{I}$ . The time origin is chosen to be the start of the 1st pulse. At $t = P$, we get $I_1 = \tilde{I}K$ . The initial





current is reduced by the multiplicative factor caused by the charges stored in the NPs during the pulse. At $t=T+P$, we get $I_2 = \tilde{I}\left(1-(1-K)e^{-(T-P)/\tau_d}\right)K$. The term between brackets is due to the discharge between the two first pulses; if $T$ is very long the NPs will be completely empty and this term will give a factor 1. The additional $K$ factor is caused by the charges trapped in the NPs during the second pulse. By doing so, the general formula (S6) can be obtained.

## VI. Simple charge/discharge experiments

An alternative measurement has been performed on the NOMFET to get more information about the discharging phenomena. We first applied a negative bias on the gate (of the same value as those used in the spike experiments shown in Fig. 6, i.e. -30V) during a sufficiently long time to be close to the maximal charge of the NPs (defined as the time for which the drain current reaches a steady state). Then, we switched the gate to 0V and we measured the time-dependent variation of the drain current corresponding to the discharge of the NPs. We fitted the discharging curves with a linear combination of exponentials

$$f_{Rd}(t) = \sum_i a_{Rid} e^{-t/\tau_{Rid}}$$

(S10)

with $\tau_{R1d} << \tau_{R2d} <<...$an so forth.

Two typical examples are shown in Fig. S2 that gives typical values for the discharge $\tau_{R1d} = 3.7$ s and $\tau_{R2d} = 19.1$s (for the NOMFET with the 5 nm NPs and L= 12 µm as on figure 6), and $\tau_{R1d} = 0.4$ s and $\tau_{R2d} = 2.2$ s (for the NOMFET with the 5 nm NPs and L= 2 µm). The fitting of the discharge current is not significantly improved by using three exponential terms (we get $R^2 > 0.999$ with two exponentials). Comparing (Fig. 7, main text) these time constants with those obtained from the fit of the iterative models on the spike experiments (Fig. 6), we find that the second relaxation time constant $\tau_{R2d}$ is very similar to $\tau_d$ for all the NOMFET with various channel lengths L and NPs sizes. It means that $\tau_{R1d}$ is too short to lay any role in the spiking experiments done here. Even if the response of the device to a sequence of pulses should be improve by adding more terms in the discharging function in equations (S6), (S9) and (3 main text),we found that the contribution of the shortest time is not





necessary to reproduce the general behavior of the NOMFET to the sequences of pulses considered in this work.

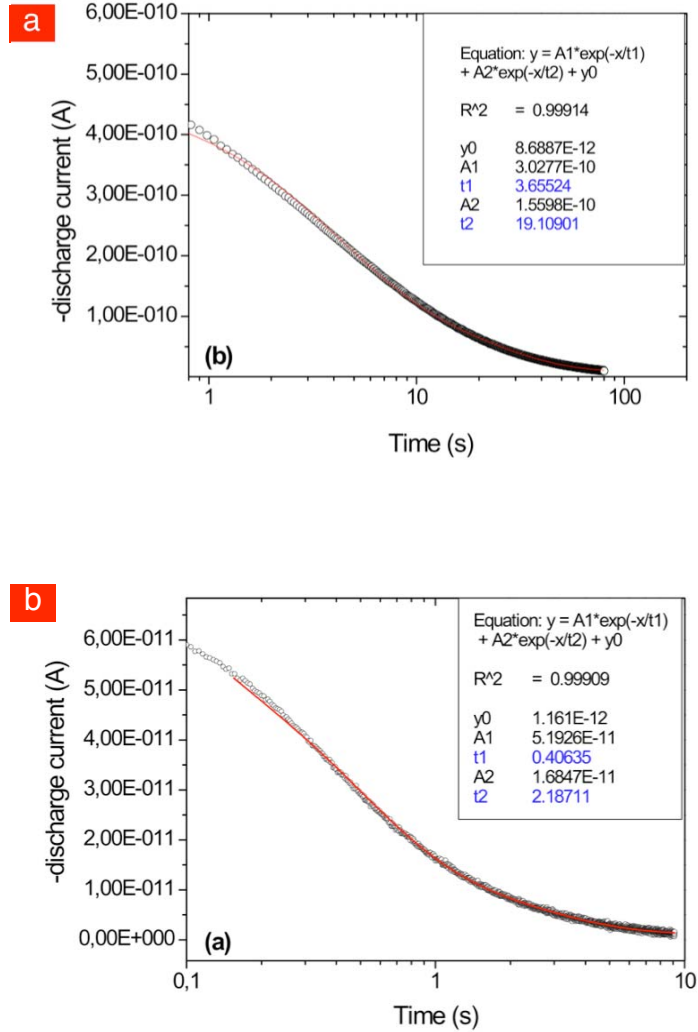

**Figure S5. a**, semi-log plot of the time dependent variation of the NOMFET (L=12 µm and 5 nm NPs) drain current during the discharge of the NPs (all terminals grounded). The red lines are the fit with equation (S10) given $\tau_{R1d}$ = 3.7 s and $\tau_{R2d}$ = 19.1s. **b**, same experiments for the NOMFET L=2 µm and 5 nm NPs. The fit gives $\tau_{R1d}$ = 0.4 s and $\tau_{R2d}$ = 2.2 s.

## VII. Amplitude of the STP versus the frequency

We observed that the amplitude of the depressing and facilitating behavior during the STP experiments (Fig. 6) depends on the frequency of the signal. For instance, the decrease in the $I_{DS}$ current is more important at 2 Hz than at 0.5 Hz (fig. 6-a). This is expected because at 2 Hz, the time





interval between two subsequent pulses is shorter than a t 0.5 Hz, and the charges stored in the NPs during the pulse period have less time to relax. At a consequence, at the end of the train of pulses, more charges remain stored in the NPs and the decrease in the current is also more important. The amplitude of the depressing behavior increases with the frequency of the pulses. The model catches this behavior satisfactorily.

Let us consider the limit for a large number of pulses, where $I_n$ reaches a steady state (Fig. S6)

$$I_n \xrightarrow{n \to +\infty} \tilde{I}\left(e^{(T-P)/\tau_d} - 1\right)\frac{Ke^{-(T-P)/\tau_d}}{1 - Ke^{-(T-P)/\tau_d}} \tag{S11}$$

In the particular case where $T << \tau_d$ (depressing behavior) this limit becomes

$$\lim_{n \to \infty} I_n = \frac{1}{1 + \dfrac{\tau_d}{T - P}\left(K^{-1} - 1\right)} \approx \frac{1}{f} \tag{S12}$$

with $f = 1/T$. This particular behavior is indeed observed experimentally (Fig. S6). This 1/f behavior was also observed by Tsodyks et al.[7] and Abbott et al.[6] in the case of a biological synapse (for instance, see figure 1 in Ref. [6]).





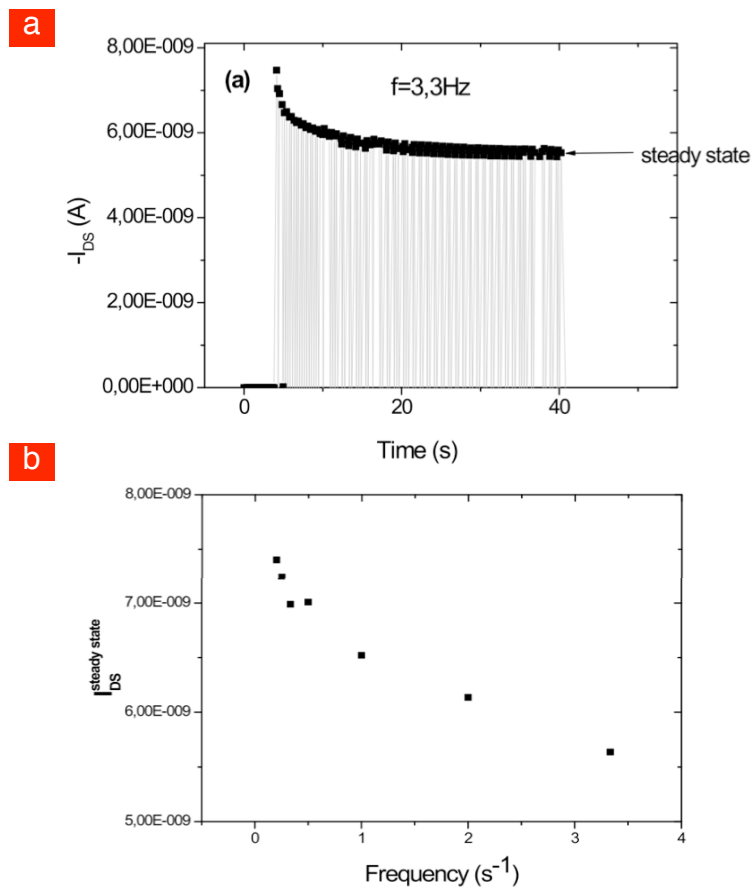

**Figure S6. a,** response of the NOMFET for a long sequence of pulses. The output current reaches a steady-state. **b**, evolution of the steady-state current as a function of the frequency of the pulses. Data for the NOMFET with the 5 nm NPs and L= 12 µm.